\documentclass[aps,prl,floatfix,showpacs,reprint]{revtex4-1}

\usepackage{graphicx}
\usepackage{dcolumn}
\usepackage{bm}
\usepackage{amssymb}
\usepackage{rotating} 
\usepackage{amsmath}

\newcommand{\sub}[1]{\ensuremath{_{\textrm{#1}}}}
\newcommand{\one}{\:^1}
\newcommand{\two}{\:^2}
\DeclareMathOperator{\Tr}{Tr}
\DeclareMathOperator{\rank}{rank}

\begin{document}

\title{Global Solutions of Hartree-Fock Theory and their Consequences for Strongly Correlated Quantum Systems}

\author{Srikant Veeraraghavan and David A. Mazziotti}

\email{damazz@uchicago.edu}

\affiliation{Department of Chemistry and The James Franck Institute,
The University of Chicago, Chicago, IL 60637}

\date{Submitted May 30, 2013; Revised January 7, 2014}

\begin{abstract}

We present a density matrix approach for computing global solutions
of Hartree-Fock theory, based on semidefinite programming (SDP),
that gives upper and lower bounds on the Hartree-Fock energy of
quantum systems.  Equality of the upper- and lower-bound energies
guarantees that the computed solution is the globally optimal
solution of Hartree-Fock theory.   For strongly correlated systems
the SDP approach provides an alternative to the locally optimized
Hartree-Fock energies and densities from the standard solution of
the Euler-Lagrange equations. Applications are made to the potential
energy curves of the H$_{4}$ dimer and the N$_{2}$ molecule.

\end{abstract}


\maketitle

In the 1950s Roothan \cite{R51} developed the self-consistent-field
approach to solving Hartree-Fock theory.  In addition to providing
some of the earliest electronic structure calculations of atoms and
molecules, Roothan's method with a variety of refinements to
accelerate convergence~\cite{SH73, P80, C00, T04, FMM04, GS12} and
to treat large systems~\cite{McWeeny56, LNV93, S96, Y97, M97, PM98,
G99,  M01, N02, JBO13} has remained the standard approach to
implementing Hartree-Fock theory for nearly 60~years.  In this
Letter we present a density matrix approach to Hartree-Fock theory,
based on semidefinite programming (SDP)~\cite{SDP96, BM05, M04,
*M11}, that gives upper and lower bounds on the Hartree-Fock energy
of atoms and molecules.  For strongly correlated quantum systems
this SDP approach provides an alternative to the locally optimized
Hartree-Fock energies and densities from standard algorithms.

Two complementary semidefinite programming formulations of
Hartree-Fock theory are derived that yield upper and lower bounds on
the Hartree-Fock energy, respectively.  When the upper- and
lower-bound energies are equal, we have a guaranteed certificate
that the computed solution is the global solution of Hartree-Fock
theory. The computed energy is the global energy minimum, and the
computed density is the electron density at that global minimum. We
illustrate the methodology through applications to the dissociation
curves of the H$_{4}$ dimer and the nitrogen molecule.  In both
examples, we compare SDP and DIIS algorithms that implement the {\em
restricted} Hartree-Fock method in which the spin symmetry is
conserved~\cite{Jensen}.  At geometries away from equilibrium, the
SDP Hartree-Fock method yields spatial-symmetry-broken Hartree-Fock
energies that are lower than those from the standard DIIS
Hartree-Fock method. When the electron correlation is significant,
the standard DIIS Hartree-Fock method or other methods based on the
solution of the Euler-Lagrange equations have difficulty selecting
the global minimum from multiple local minima of different spatial
symmetries. These solutions from SDP satisfy the standard
Euler-Lagrange equations just like the higher-in-energy DIIS
solutions.

The self-consistent-field Hartree-Fock method for an $N$-electron
system iteratively solves a system of Euler-Lagrange equations for a
stationary point.  The stationary point yields a ground-state
Hartree-Fock energy and a set of $N$ occupied orbitals.  The
computed Hartree-Fock energy is not guaranteed to be the global
energy minimum.  From the perspective of reduced density matrices
(RDMs)~\cite{RDM07,CB}, we can understand the self-consistent-field
method as iteratively checking extreme points of the set of
one-electron RDMs (1-RDMs) for satisfaction of the Euler-Lagrange
equations where each extreme point corresponds to a 1-RDM with a
Slater determinant preimage~\cite{Nrep,H78,Nrep2}. The optimization
of the Hartree-Fock energy over the set of extreme 1-RDMs (those
with an $N$-electron Slater determinant as a preimage) can be
replaced without approximation by an optimization over the larger
(and convex) set of $N$-representable 1-RDMs (those with any
$N$-electron wave function as a preimage)~\cite{Lieb81, C00}:
\begin{align}
\operatorname*{minimize}_{\one D, \one Q \in \mathbb{H}^r_+}
& \hspace{0.25cm} E_{\rm HF}(\one D) \label{eq:rdm} \\
\operatorname*{subject\,to}
& \Tr(\one D) = N \label{d1tr} \\
& \one D + \one Q = I \label{d1q1}
\end{align}
where $E_{\rm HF}$ is the following quadratic function of the 1-RDM:
\begin{align}
E_{\rm HF}(\one D) & = \sum_{ij}^r \one K^i_j \one D^i_j +
\sum_{ijkl}^r \one D^i_k \two
V^{ik}_{jl} \one D^j_l \\
\one K^i_j & = \langle i| {\hat h} |j \rangle \\
\two V^{ik}_{jl} & =  \frac{1}{2}(\langle ij|kl \rangle - \langle
ij|lk \rangle). \label{eq:v2}
\end{align}
The one-electron Hamiltonian operator ${\hat h}$ contains the
kinetic energy operator and electron-nuclei potential, $\langle
ij|kl \rangle$ represents the electron-electron repulsion integrals,
and the indices $i$, $j$, $k$, and $l$ denote the orbitals in the
one-electron basis set of rank $r$. The notation $\one D, \one Q \in
\mathbb{H}^r_+$, equivalent to $\one D \succeq 0$ and $\one Q
\succeq 0$, indicates that both the 1-particle RDM $\one D$ and the
1-hole RDM $\one Q$ are contained in the set of $r \times r$
Hermitian positive semidefinite matrices.

The reduced-density-matrix formulation of Hartree-Fock theory can be
recast as a convex semidefinite program by embedding the quadratic
product of 1-RDMs in $E_{\rm HF}$ in a higher dimensional
(two-electron) matrix $\two M \in \mathbb{H}^{r^2}_+$. Rewriting
$E_{\rm HF}$ as a linear functional of $\two M$
\begin{align}
 E(\one D, \two M) & = \Tr(\one K \one D) + \Tr(\two V \two M),
\end{align}
we can relax the non-convex Hartree-Fock optimization to a convex
semidefinite program:
\begin{align}
\operatorname*{minimize}_{\one D, \one Q \in \mathbb{H}^{r}_+, \two
M \in \mathbb{H}^{r^2}_+}
 & E(\one D, \two M) \label{eq:minM} \\
\operatorname*{subject\,to} \hspace{0.5cm}
& \Tr(\one D) = N  \\
& \Tr(\two M) \le N  \\
& \one D + \one Q = I  \\
& \sum_{j=1}^r \two M^{ik}_{jj} = N \one D^i_k . \label{eq:conM}
\end{align}
The solution of this SDP relaxation yields a lower bound to the
Hartree-Fock energy.  Because the constraints on the matrix $\two M$
are minimal, this convex SDP formulation will typically yield
energies that are significantly below the Hartree-Fock energy. To
reproduce Hartree-Fock, further constraints on $\two M$ are
required.

\begin{figure}[htp!]

\includegraphics[scale=0.7]{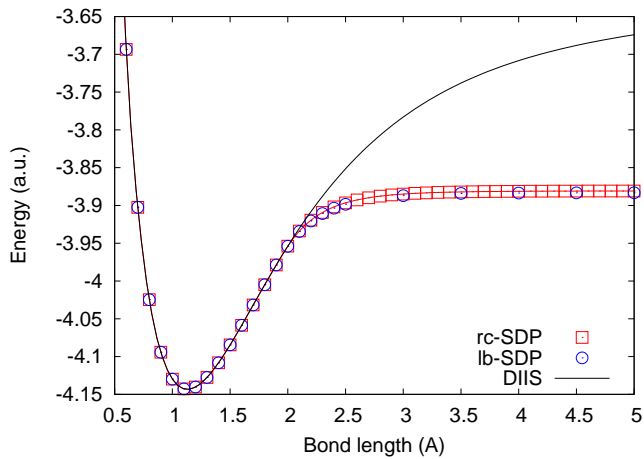}

\caption{Upper and lower bounds to the ground-state restricted
Hartree-Fock energy from rc-SDP and lb-SDP, respectively, as well as
the Hartree-Fock energy from DIIS are shown as functions of the
distance between the H\sub4 monomers.  The rc-SDP energy is
certified by lb-SDP to be globally optimal for $R \le 2$~\AA\ and
within 0.003~a.u. of the globally optimal solution for all $R$.  In
contrast, the DIIS solutions converge to local solutions for all
separations $R \ge 2$~\AA.}

\label{H8-HF}

\end{figure}

Two separate sets of additional conditions on the matrix $\two M$
that yield upper and lower bounds on the Hartree-Fock energy,
respectively, will be considered.  The first set of constraints,
yielding the upper bound, consists of a single rank constraint
\begin{equation}
\label{eq:rr} \rank{\left(\two M\right)} = 1 .
\end{equation}
The $\two M \in \mathbb{H}^{r^2}_+$ matrix with its rank-one
constraint and the contraction constraint in Eq.~(\ref{eq:conM}), we
can show, is a tensor product of two identical 1-RDMs
\begin{align}
\label{eq:mrr} \two M^{ik}_{jl} & = \one D^i_k \one D^j_l .
\end{align}
It follows that the solution of the optimization program in
Eqs.~(\ref{eq:minM}-\ref{eq:conM}) with the rank constraint in
Eq.~(\ref{eq:rr}) is equivalent to the solution of the RDM
formulation of Hartree-Fock theory in
Eqs.~(\ref{eq:rdm}-\ref{eq:v2}).  We have mapped Hartree-Fock theory
exactly onto a rank constrained semidefinite program (rc-SDP
HF)~\cite{DB}. The rank constrained semidefinite program is convex
except for the rank restriction; the nonconvexity of the
Hartree-Fock energy functional in the RDM formulation has been
transferred to the rank restriction in the SDP formulation. Because
of the rank constraint, the solution of rc-SDP HF is not necessarily
a global solution, meaning that the solution can be a local minimum
in the Hartree-Fock energy and hence, an upper bound on the global
energy minimum. Unlike traditional formulations of Hartree-Fock
theory, however, rc-SDP HF optimizes the 1-RDM over the convex set
of $N$-representable 1-RDMs, and in practice, we find that this
difference makes it much more robust than traditional formulations
in locating the global solution.

The second set of conditions, yielding a lower bound, consists of
four constraints including
\begin{equation}
\sum_{j=1}^r \two M^{ij}_{jk} = \one D^i_k
\end{equation}
and three additional constraints from permuting the indices $i$ and
$j$ and/or $j$ and $k$ symmetrically.  These convex conditions are a
relaxation of the idempotency of the 1-RDM. They are necessary but
not sufficient for the idempotency of the 1-RDM at the Hartree-Fock
solution, and hence, optimization of the SDP program in
Eqs.~(\ref{eq:minM}-\ref{eq:conM}) with these additional constraints
(lb-SDP) is an SDP relaxation of the reduced-density-matrix
formulation of Hartree-Fock theory in
Eqs.~(\ref{eq:rdm}-\ref{eq:v2}). The lb-SDP method yields a {\em
lower bound on the energy} from the global Hartree-Fock solution. In
practice, this lower bound is found to be quite tight, and in some
cases it agrees exactly with the global Hartree-Fock solution. If
lb-SDP produces a 1-RDM solution that is idempotent, then that
solution is the global Hartree-Fock solution. Furthermore, when the
upper and lower bounds from rc-SDP and lb-SDP agree, we have a
guaranteed certificate that these computed bounds correspond to the
global energy minimum of Hartree-Fock theory.

To illustrate the rc-SDP and lb-SDP methods, we apply them to
computing the restricted Hartree-Fock dissociation curves for
singlet (H\sub4)\sub2 and N\sub2 in the cc-pVDZ basis
set~\cite{D89}. The GAMESS electronic structure package is used to
perform self-consistent-field Hartree-Fock calculations (SCF HF with
DIIS) and coupled cluster singles-doubles (CCSD)~\cite{PB82,BM07,
CCSD} calculations. The rc-SDP and lb-SDP are solved using the SDP
solver RRSDP~\cite{M04}. Since DIIS is the standard accelerator for
SCF HF calculations, we compare rc-SDP HF results with DIIS results.
Both rc-SDP HF and DIIS methods are performed without enforcing a
specific spatial symmetry.  The DIIS solution at the internuclear
distance $R'$ where $R'$ is differentially larger than the distance
$R$ is obtained by using the DIIS solution at $R$ as an initial
guess.

The SDP solver RRSDP imposes the semidefinite constraint on each
matrix $M$ through the factorization $M = RR^T$.  For rc-SDP HF, the
rank-one constraint on $\two M$ is readily enforced by defining $R$
to be a rectangular $r \times 1$ matrix. Scaling of RRSDP \cite{M04}
is determined by the $RR^T$ matrix multiplication for the largest
matrix block, which is the $r^{2} \times r^{2}$ matrix $\two M$ for
both rc-SDP and lb-SDP. For rc-SDP the rank of $\two M$ is one, and
hence, the matrix multiplication scales approximately as $r^4$. For
lb-SDP the rank of $\two M$ scales as $r$ after applying the bound
on the maximum rank from Pataki~\cite{P98} and Barvinok~\cite{B95},
and hence, the matrix multiplication scales approximately as $r^5$.

\begin{figure}[htp!]

\includegraphics[scale=0.7]{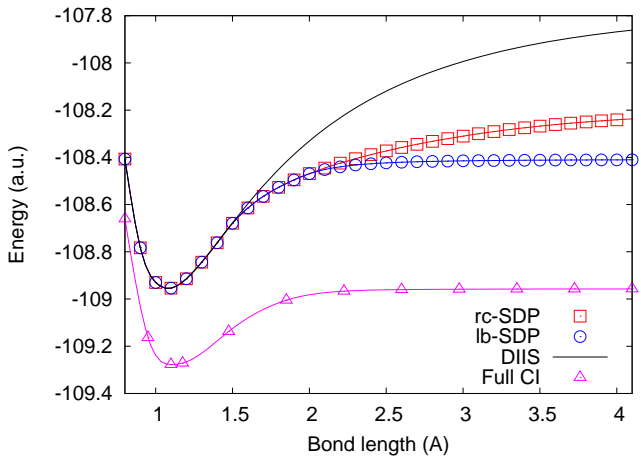}

\caption{The ground-state restricted Hartree-Fock energies from
rc-SDP and DIIS and the lower bound from lb-SDP are shown as
functions of the N-N internuclear distance $R$. When the energies
from rc-SDP and lb-SDP agree, the solution from rc-SDP is guaranteed
to be the global solution of Hartree-Fock theory.  At stretched
geometries the DIIS locates a D$_{4{\rm h}}$ solution while rc-SDP
locates an energetically lower pair of C$_{2{\rm v}}$ solutions. By
4.1~\AA\ the D$_{4{\rm h}}$ and C$_{2{\rm v}}$ solutions differ by
0.375~a.u. (235~kcal/mol). The C$_{2{\rm v}}$ solutions can be
considered excited states in the Hartree-Fock model since they
analytically continue to excited-state solutions at equilibrium
geometries (not shown); however, because the two iso-energetic
C$_{2{\rm v}}$ Hartree-Fock solutions are also the global minimum at
stretched geometries, they provide insight into the correlated
ground-state solution of the Schr{\"o}dinger equation.}

\label{N2-HF}

\end{figure}

H\sub4 is known to be a multireferenced diradical system at square
geometries~\cite{FCF95, SandH4}. Hence, molecules constructed from
square H\sub4's can be expected to be challenging systems for
Hartree-Fock and correlation methods.  Here we examine the dimer of
square H\sub4 molecules whose H-H bond lengths are 1.0~\AA.
Figure~\ref{H8-HF} shows the upper and lower bounds to the
ground-state restricted Hartree-Fock energy from rc-SDP and lb-SDP,
respectively, as well as the restricted Hartree-Fock energy from
DIIS as functions of the distance $R$ between the H\sub4 monomers.
Because the lb-SDP curve is identical to the rc-SDP and DIIS curves
for $R \le 2.0$~\AA\, we have a certificate of global optimality for
that part of the curve. Furthermore, since lb-SDP is never lower
than rc-SDP by more than 0.003~a.u. after 2.0~\AA\, the rest of the
rc-SDP curve is equal to the globally optimal Hartree-Fock solution
within that threshold. Its optimality is corroborated by the fact
that the rc-SDP curve is size consistent, meaning that it is
asymptotically equal to exactly twice the restricted Hartree-Fock
energy of the singlet H\sub4 monomer.  The energy of the monomer is
the same from both rc-SDP and DIIS and certified to be globally
optimal within 0.001 a.u. by lb-SDP.

For all separations of the dimer larger than 2.0 ~\AA\, the DIIS
method converges to a local solution with D$_{4{\rm h}}$ symmetry
while the rc-SDP method converges to the global C$_{2{\rm v}}$
solution.  At $R = 5.0$~\AA\ the DIIS energy of the dimer is more
than 0.2~a.u. (125~kcal/mol or the energy of a chemical bond) higher
than the size consistent energy of the global Hartree-Fock solution.
If the DIIS algorithm is seeded with an rc-SDP solution, it
converges directly to that solution, showing that like the local
DIIS solutions the rc-SDP solutions also satisfy the restricted
Euler-Lagrange equations.  In principle, even from a poor initial
guess, the C$_{2{\rm v}}$ solution can also be obtained from DIIS
from a repeated application of stability analysis involving the
potentially expensive construction of the Hessian matrix~\cite{CP67,
SP77}.  Unlike the SDP methods, however, the DIIS with stability
analysis cannot determine whether a local solution is the global
solution.  Both the DIIS and the rc-SDP restricted Hartree-Fock
methods generate solutions with $\langle {\hat S}^2 \rangle = 0$ for
all $R$ while the unrestricted Hartree-Fock method produces a
solution with $\langle {\hat S}^2 \rangle = 2$ for $R > 2.25$~\AA.

An accurate description of a stretched triple bond in N\sub2 is a
challenging electronic structure problem~\cite{Piecuch01}.
Figure~\ref{N2-HF} shows the ground-state restricted Hartree-Fock
energies from rc-SDP and DIIS and the lower bound from lb-SDP as
functions of the N-N internuclear distance $R$. Agreement of the
energies from rc-SDP and lb-SDP guarantees that the solution from
rc-SDP is the global solution of Hartree-Fock theory until
$R=2.0$~\AA.  After 2.0~\AA\ lb-SDP produces a non-idempotent 1-RDM
and gives a lower bound on the Hartree-Fock energy; in this
stretched region the rc-SDP likely continues to give the globally
optimal curve although we do not have a formal mathematical
guarantee.   As in the (H\sub4)\sub2 example, after 1.5~\AA\ the
DIIS potential energy curve smoothly diverges from the rc-SDP global
solution to a local solution.  At 4.1~\AA\ the rc-SDP energy per N
atom is $-54.118$~a.u. which is 0.187~a.u. lower than the energy per
N atom from DIIS and 0.270~a.u. higher than the energy of a single
nitrogen in its quadruplet state from the restricted open-shell
Hartree-Fock method.

For all internuclear distances larger than 1.6~\AA\, the DIIS method
converges to a local solution with D$_{4{\rm h}}$ symmetry while the
rc-SDP method converges to the global C$_{2{\rm v}}$ solution.  We
can interpret the N$_{2}$ solution from rc-SDP as an ensemble
mean-field density matrix with D$_{4{\rm h}}$ symmetry, composed of
two iso-energetic C$_{2{\rm v}}$ Slater determinants.  As in the
previous example, the rc-SDP solutions were additionally verified to
be HF minima by showing that as initial guesses for the DIIS
algorithm they satisfy the Euler-Lagrange equation.  In principle,
the C$_{2{\rm v}}$ determinants from rc-SDP can be computed with
DIIS in combination with Hessian-based stability analysis; unlike
the SDP approach, however, stability analysis cannot determine
whether a local solution is also a global solution~\cite{CP67,SP77}.
Unlike the unrestricted Hartree-Fock solution, where $\langle {\hat
S}^2 \rangle = 3$ after $R = 2.0$~\AA, the rc-SDP and lb-SDP
solutions have $\langle {\hat S}^2 \rangle$ identically equal to
zero for all $R$.

\begin{figure}[htp!]

\includegraphics[scale=0.7]{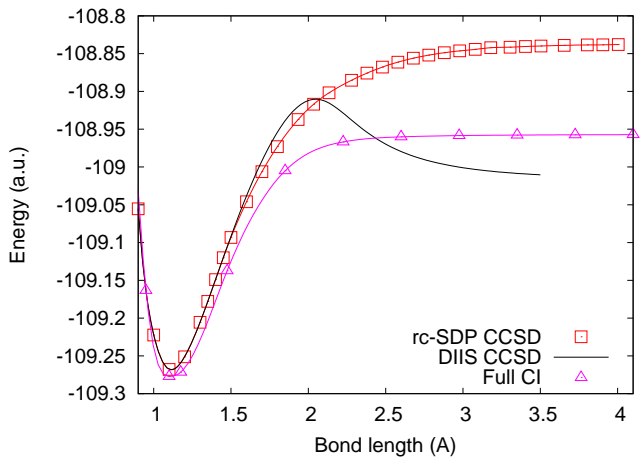}

\caption{The potential energy curve of N$_{2}$ from CCSD performed
with the rc-SDP restricted Hartree-Fock reference wave function is
compared with the potential energy curve from CCSD performed with
the standard DIIS restricted Hartree-Fock reference wave function.
The unphysical nature of the CCSD curve, well documented in the
literature, has been attributed to missing electron correlation.  As
the bond is stretched, while CCSD with the D$_{4{\rm h}}$ reference
from DIIS rises rapidly and then diverges, CCSD with the C$_{2{\rm
v}}$ reference from rc-SDP yields a physically realistic
dissociation curve.}

\label{N2-CCSD}

\end{figure}

Changes in Hartree-Fock energies and densities can impact
correlation energy calculations in two ways: (1) any change in the
Hartree-Fock energy changes the correlation energy by its very
definition and (2) any change in the Hartree-Fock density (or the
Hilbert space spanned by the molecular orbitals) changes the
reference wave function employed in many-electron correlation
methods including coupled cluster~\cite{PB82,BM07,CCSD} and
parametric RDM methods~\cite{M08,*M10}. Figure~\ref{N2-CCSD}
explores the effect of using the global C$_{2{\rm v}}$ solution
rather than the D$_{4{\rm h}}$ solution as the reference wave
function in CCSD. The treatment of N$_{2}$ by CCSD has been widely
documented in the literature where the unphysical behavior of the
N$_{2}$ curve from CCSD has been attributed to the absence of $T_3$
and $T_4$ excitation amplitudes~\cite{bartlett87,M04}. The results
in Fig.~\ref{N2-CCSD}, however, show that in the dissociation limit,
while CCSD with the D$_{4{\rm h}}$ reference rises rapidly and then
diverges, CCSD with the C$_{2{\rm v}}$ reference yields a physically
realistic dissociation curve.

State-of-the-art calculations on the H$_{4}$ dimer~\cite{FCF95,N02b,
SandH4} and the N$_{2}$ molecule~\cite{ACSEN2} require an explicit
treatment of multiple determinants in the reference wavefunction,
known as multireference correlation.  The presented global-minimum
Hartree-Fock theory can be useful in identifying the determinants in
the H$_{4}$ dimer and the N$_{2}$ molecule that contribute most
significantly to the strong electron correlation. Symmetry-broken
determinants, for example, can be employed as an $N$-electron basis
in a symmetry-restoring configuration interaction.

The SDP-based Hartree-Fock algorithms have also been applied to
Cr\sub2, CN, and NO\sub2~\cite{JCP} where there are many local
solutions, especially in the case of Cr\sub2. In these and other
larger systems with symmetry breaking, alternatives to the local
solution of the Hartree-Fock equations become essential to
identifying the global solution. Because stability analysis only
distinguishes local minima from saddle points~\cite{CP67, SP77}, it
becomes less useful in the presence of multiple local solutions as
can occur in larger strongly correlated molecular systems and
materials.

We have presented an RDM formulation of Hartree-Fock theory, based
on semidefinite programming, that yields upper and lower bounds on
the Hartree-Fock solution.  When these bounds are equal, they
provide a certificate guaranteeing the globally optimal Hartree-Fock
solution. As electrons become more strongly correlated, methods for
Hartree-Fock based on the self-consistent-field approach like DIIS
can converge to stationary points of Hartree-Fock theory with
potentially non-global energies and densities.  The SDP-based
restricted Hartree-Fock method is directly extendable to a
restricted open-shell Hartree-Fock method and an unrestricted
Hartree-Fock method, which will be presented elsewhere.

\begin{acknowledgments}

DAM gratefully acknowledges the NSF, ARO, and Microsoft Corporation
for their generous support.

\end{acknowledgments}

\bibliography{hf_sdp_r3}

\end{document}